\documentclass[a4paper,12pt]{article}

\usepackage[latin1]{inputenc}
\usepackage[top=2.5cm, left=2.5cm, right=2cm, bottom=2.5cm]{geometry}
\usepackage{graphicx}
\usepackage[centertags]{amsmath}
\usepackage{amsfonts}
\usepackage{amssymb}
\usepackage{amsthm}
\usepackage{newlfont}
\usepackage{xcolor}
\usepackage{soul}

\begin{document}
%\half
\begin{center}

{\Large \bf{New boundary conditions for (extended) $\mathrm{AdS}_3$ supergravity}}

\vspace{1cm}

C. E. Valc\'arcel\footnote{valcarcel.flores@gmail.com}

\vspace{.5cm}

$^{a}$\emph{Instituto de F\'isica, Universidade Federal da Bahia, C\^ampus Universit\'ario de Ondina, 40210-340, Salvador-BA, Brazil}.\\
$^{b}$\emph{CMCC-Universidade Federal do ABC, Santo Andr\'e, SP, Brazil}.

\end{center}

\vspace{.25cm}

\begin{abstract}

In this work, we build the most general boundary conditions for $\mathcal N=\left(1,1\right)$ and $\mathcal N=\left(2,2\right)$ extended supergravity. We show that for the loosest set of boundary conditions, their asymptotic symmetry algebras are two copies of the $\mathfrak{osp}\left(1\mid 2\right)_k$ and $\mathfrak{osp}\left(2\mid 2\right)_k$  algebra, respectively. Then, we restrict the gauge fields on the boundary conditions in order to obtain supersymmetric extensions of the Brown-Henneaux and Troessaert boundary conditions. Furthermore, for $\mathcal N=1$ and $\mathcal N=2$ (extended) supergravity we show the supersymmetric extensions of Avery-Poojary-Suryanarayana boundary conditions.

\vspace{.5cm}


\end{abstract}

\section{Introduction}

In their seminal work \cite{BH}, Brown and Henneaux showed that the asymptotic symmetries of $\mathrm{AdS_3}$ gravity are two copies of the Virasoro algebra. This result is known as a precursor of the $\mathrm{AdS/CFT}$ correspondence \cite{Maldacena} and a realization of the Holographic Principle \cite{Susskind}. Since then, there has been several extension of $\mathrm{AdS_3}$ holography. For instance, in \cite{Banados-SUSY} and \cite{Henneaux-SUSY} were investigated the asymptotic symmetries of supergravity and extended supergravity, respectively. These works demonstrated that the asymptotic symmetry algebra for (extended) supergravity is the (extended) super-conformal algebra \cite{superconformal}, \cite{extended-susy}, \cite{extended-susy2}.

The study of asymptotic symmetries of diverse models of three-dimensional supergravity still draws a lot of attention since it might sheds light on the geometrical origin of superconformal algebras. We can cite, for example, the asymptotic analysis of supersymmetric higher-spin gravity \cite{SuperW}, \cite{Tan}, \cite{Peng}, hypergravity \cite{Hyper}, \cite{Hyper2} and supergravity in the flat-space limit \cite{flat1}, \cite{flat2}, \cite{flat3}, \cite{flat4}, \cite{flat5}, \cite{flat6}. An important tool to obtain those results is the Chern-Simons formulation of three-dimensional gravity \cite{Witten}, which allows a simpler analysis of these symmetries compared to the metric formulation (see  \cite{note01} and \cite{note02} for pedagogical reviews).

An important aspect of $\mathrm{AdS_3}$ holography is the choice of fall-offs conditions for the metric. For pure $\mathrm{AdS_3}$ gravity, the most common ones are the Brown-Henneaux boundary conditions \cite{BH}. Nevertheless, in the last years new sets of boundary conditions have been proposed: Comp\`ere, Song, Strominger \cite{Compere} and Troessaert \cite{Troessaert} proposed, independently, new boundary conditions that led to two copies of a warped conformal algebra (with different Kac-Moody levels). Avery, Poojary and Suryanarayana \cite{Avery} proposed boundary conditions that led to a semidirect sum of an $\mathfrak{sl}\left(2\right)_k$ algebra and a Virasoro algebra. In \cite{Grumiller3D}, Grumiller and Riegler worked with the most general boundary conditions for $\mathrm{AdS_3}$ gravity, as a result, the asymptotic symmetry algebra was generated by two copies of the $\mathrm{sl}\left(2\right)_k$ algebra. In addition, all other previously found boundary conditions were obtained imposing some restrictions on the most general ones. This procedure has also been applied to three-dimensional flat-space \cite{GrumillerFlat}, chiral higher spin gravity \cite{Krishnan} and in the context of (supersymmetric) two-dimensional gravity \cite{AdS2}, \cite{AdS2SUSY}. The aim of this paper is to extend the analysis of \cite{Grumiller3D} to three-dimensional $\mathrm{AdS}$ supergravity: We obtain the asymptotic symmetry algebra for the loosest set of boundary conditions for (extended) supergravity, reproduce some old results and also show new boundary conditions related to warped super-conformal algebras. Therefore, this procedure provides a good laboratory to explore the rich asymptotic structure of (extended) supergravity.

This paper is organized as follows. In section \ref{sec2}, we give a brief review of the Chern-Simons formulation for $\mathcal N=(N,M)$ supergravity. In section \ref{sec3}, we introduce the most general boundary conditions for $\mathcal N=\left(1,1\right)$ supergravity and in section \ref{sec4} we present three particular cases: The first two are supersymmetric extensions of the Brown-Henneaux and Troessaert boundary conditions, the third one is the supersymmetric Avery-Poojary-Suryanarayana boundary conditions for $\mathcal N=1$ supergravity. In section \ref{sec5}, we work with the most general boundary conditions for $\mathcal N=\left(2,2\right)$ extended supergravity and in section \ref{sec6} we focus on some particular boundary conditions. Finally, in section \ref{sec7} we comment our results and give future perspectives of this work.

\section{Chern-Simons formulation of $\mathcal N=(N,M)$ $\mathrm{AdS_3}$ supergravity}\label{sec2}

It is well known that the three-dimensional $\mathcal{N}=(N,M)$ supergravity with negative cosmological constant can be described as the difference of two Chern-Simons action under the $\mathrm{OSp}\left(N\mid 2\right)\otimes \mathrm{OSp}\left(M \mid 2\right)$ group\footnote{For $N,M$ greater or equal than $2$, the theory is called extended supergravity} \cite{Achucarro}: \begin{equation}\label{sugra01}
    \mathrm{I_{Sugra}}=\mathrm{I_{CS}}\left[ \Gamma \right] - \mathrm{I_{CS}}\left[ \bar{\Gamma} \right],
\end{equation}
where
\begin{equation}\label{sugra02}
    \mathrm{I_{CS}}\left[ \Gamma \right] = \frac{k}{4\pi}\int_{\mathcal M} \mathrm{Str}\left[ \Gamma \wedge \mathrm{d}\Gamma + \frac{2}{3} \Gamma\wedge \Gamma\wedge \Gamma\right] + \mathrm{B} \left[ \Gamma \right].
\end{equation}
In the above equation $k$ is the Chern-Simons level, which is related to the Newton constant $G$ and the $\mathrm{AdS}$ radius $l$ by $k=l/4G$. The connection $1-$ forms $\Gamma,\;\bar{\Gamma}$ are valued on the $\mathfrak{osp}\left(N\mid  2\right)$ and $\mathfrak{osp}\left(M\mid 2\right)$ super-algebra, respectively (See Appendix). The integration in the Chern-Simons action $(\ref{sugra02})$ is performed over a manifold $\mathcal M$ which has the topology of a cylinder. The radial coordinate is denoted by $\rho$ and the angular and time coordinates by $\varphi$ and $t$, respectively. Furthermore, the boundary coordinates are periodic: $\varphi \sim \varphi + 2\pi$, $t \sim t + 2\pi$. The last term in \eqref{sugra02} is the boundary term:
\begin{equation}\label{sugra03}
\mathrm{B} \left[ \Gamma \right] = \frac{k}{4\pi} \int_{\partial\mathcal M} \mathrm{d}^2x \mathrm{Str}\left[\Gamma_t \Gamma_\varphi\right]
\end{equation}
which relax the condition on the connection to $\delta \Gamma_t=0$.

Instead of working with the complete action \eqref{sugra01}, let us work with just one copy of the Chern-Simons action: $\mathrm{I_{CS}}\left[ \Gamma \right]$, which will be called $\Gamma$-sector. The action \eqref{sugra02} has equation of motion
\begin{equation}\label{sugra04}
    F = \mathrm{d}\Gamma + \Gamma \wedge \Gamma = 0
\end{equation}
and is invariant under the following gauge transformation:
\begin{equation}\label{sugra05}
    \delta_\lambda \Gamma = \mathrm{d}\lambda + \left[ \Gamma, \lambda \right]
\end{equation}
where $\lambda$ is the gauge parameter.

The gauge freedom of the Chern-Simons action allows us to write
\begin{equation}\label{sugra06}
 \Gamma	=	b^{-1}\left[\mathrm{d}+a \right]b,
\end{equation}
where $b=b\left(\rho\right)$ is the group element and $a=a\left(t,\varphi\right)$ is an auxiliary connection which depend on the boundary coordinates:
\begin{eqnarray}\label{sugra07}
a\left(t,\varphi\right) = a_{t}\left(t,\varphi\right)\mathrm{d}t+a_{\varphi}\left(t,\varphi\right)\mathrm{d}\varphi.
\end{eqnarray}
The form of the connection \eqref{sugra06} is important once that separates the radial part in the function $b$. In the study of asymptotic symmetries, the choice of $b$ is irrelevant, as long as $\delta b=0$. However, it is useful for a metric description. To understand this let us consider the ground state for $\mathrm{AdS_3}$ supergravity, which is defined for vanishing gravitini: In this case, the standard choice for the group element is $b=\exp(\rho L_0)$, where $L_0$ is a generator of the $\mathfrak{sl}\left(2\right)$ algebra. However, in \cite{Grumiller3D} is used $b=\exp(L_{-1})\exp(\rho L_0)$, which allows a more general metric
\begin{eqnarray}\label{sugra08}
    \mathrm{d}s^2 &=& \mathrm{d}\rho^2 + 2\left[ e^\rho N^{(0)}_i +  N^{(1)}_i +  e^{-\rho} N^{(2)}_i + \mathcal O \left( e^{-2\rho}\right)\right]\mathrm{d}\rho \mathrm{d}x^i\nonumber\\
    &+& \left[ e^{2\rho} g^{(0)}_{ij} +  e^\rho g^{(1)}_{ij} +  g^{(2)}_{ij} + \mathcal O \left( e^{-\rho}\right)\right]\mathrm{d}x^i \mathrm{d}x^j.
\end{eqnarray}
This is the asymptotic $\mathrm{AdS_3}$ metric. We need to choose boundary conditions for supergravity such that they preserve the form of this metric.

Regge and Teitelboim showed \cite{Regge} that in the presence of a boundary the generators of gauge symmetries have to be modified with an additional term, called canonical boundary charge $\mathbb{Q}$, in order to exhibit the asymptotic symmetry algebra. Particularly, for the Chern-Simons action the variation of the boundary charge is given by \cite{note01}, \cite{note02}:
\begin{equation}\label{sugra09}
\delta_\lambda \mathbb{Q}=\frac{k}{2\pi}\int\mathrm{d}\varphi\;\mathrm{Str}\left(\lambda \delta \Gamma_{\varphi}\right).
\end{equation}
As we will shown in the following sections, the integrability of the above expression is depends on the choice of boundary conditions and the specific form of the gauge parameter.

\section{Most general boundary conditions for $\mathcal N=\left(1,1\right)$ supergravity}\label{sec3}

In this section we will work with the most general boundary condition for $\mathcal N=\left(1,1\right)$ supergravity. We begin writing the auxiliary connection $a_\varphi$ in the $\mathfrak{osp}\left(1\mid 2\right)$ basis:
\begin{eqnarray}\label{bc01}
a_\varphi\left(t,\varphi\right) &=&
-\frac{2\pi}{k}
\left[\mathcal{L}^{+}L_{1}-2\mathcal L_{0}+\mathcal{L}^{-}L_{-1}
-\frac{1}{2}\mathcal{Q}^{\frac{1}{2}}G_{\frac{1}{2}}+\frac{1}{2}\mathcal{Q}^{-\frac{1}{2}}G_{-\frac{1}{2}}\right].
\end{eqnarray}
Note that we have five state-dependent functions: three bosonic $\mathcal L^I$ and two fermionic $\mathcal{Q}^\alpha$.
For the auxiliarly connection $a_t$, we use the following boundary condition:
\begin{eqnarray}\label{bc02}
a_{t}\left(t,\varphi\right)	&=&	 \mu^{+}L_{1}+\mu^{0}L_{0}+\mu^{-}L_{-1}+\nu^{\frac{1}{2}}G_{\frac{1}{2}}+\nu^{-\frac{1}{2}}G_{-\frac{1}{2}}.
\end{eqnarray}
Here we have also introduced five independent functions $\left(\mu^I,\nu^\alpha\right)$ of the boundary coordinates. However, they are not allowed to vary. They are usually called ``chemical potentials".

The flat connection condition \eqref{sugra04} now becomes:
\begin{eqnarray}
\partial_{t}\mathcal{L}^{\pm}	&=&	 -\frac{k}{2\pi}\partial_{\varphi}\mu^{\pm}\pm2\mathcal{L}^{0}\mu^{\pm}\pm\mathcal{L}^{\pm}\mu^{0}
\pm\mathcal{Q}^{\pm\frac{1}{2}}\nu^{\pm\frac{1}{2}}\label{bc03a}\\
\partial_{t}\mathcal{L}^{0}	&=&	 \frac{k}{4\pi}\partial_{\varphi}\mu^{0}-\mathcal{L}^{+}\mu^{-}+\mathcal{L}^{-}\mu^{+}
-\frac{1}{2}\mathcal{Q}^{\frac{1}{2}}\nu^{-\frac{1}{2}}+\frac{1}{2}\mathcal{Q}^{-\frac{1}{2}}\nu^{\frac{1}{2}}\label{bc03b}\\
\partial_{t}\mathcal{Q}^{\pm\frac{1}{2}} &=& \pm\frac{k}{\pi}\partial_{\varphi}\nu^{\pm\frac{1}{2}}-2\mathcal{L}^{\pm}\nu^{\mp\frac{1}{2}}
-2\mathcal{L}^{0}\nu^{\pm\frac{1}{2}}\pm\mu^{\pm}\mathcal{Q}^{\mp\frac{1}{2}}\pm\frac{1}{2}\mu^{0}\mathcal{Q}^{\pm\frac{1}{2}}.\label{bc03c}
\end{eqnarray}
The above equations represent the temporal evolution of the state-dependent functions: $\mathcal L^I,\;\mathcal Q^\alpha$.

We now compute the gauge transformations that preserve the boundary conditions. First, we write the gauge parameter in the $\mathfrak{osp}\left(1\mid2\right)$ basis
\begin{equation}
    \lambda=b^{-1}\left[\epsilon^{+}L_{1}+\epsilon^{0}L_{0}+\epsilon^{-}L_{-1}
    +\xi^{\frac{1}{2}}G_{\frac{1}{2}}+\xi^{-\frac{1}{2}}G_{-\frac{1}{2}}\right]b\label{bc04}
\end{equation}
where $\epsilon^I$ and $\xi^\alpha$ are arbitrary bosonic and fermionic functions of the boundary coordinates. We are interested in gauge parameters that satisfy
\begin{equation}
    \mathrm{d}\lambda + \left[\Gamma,\lambda \right] = \mathcal{O}\left( \delta \Gamma\right),\label{bc05}
\end{equation}
where $\delta \Gamma$ is given by \eqref{sugra06}. The condition above impose that the transformations on the gauge components are given by:
\begin{eqnarray}
\delta_{\lambda}\mathcal{L}^{0}	&=&	 \frac{k}{4\pi}\partial_{\varphi}\epsilon^{0}-\mathcal{L}^{+}\epsilon^{-}+\mathcal{L}^{-}\epsilon^{+}
-\frac{1}{2}\mathcal{Q}^{\frac{1}{2}}\xi^{-\frac{1}{2}}+\frac{1}{2}\mathcal{Q}^{-\frac{1}{2}}\xi^{\frac{1}{2}}\label{bc06a}\\
\delta_{\lambda}\mathcal{L}^{\pm}	&=&	 -\frac{k}{2\pi}\partial_{\varphi}\epsilon^{\pm}\pm2\mathcal{L}^{0}\epsilon^{\pm}\pm\mathcal{L}^{\pm}
\epsilon^{0}\pm\mathcal{Q}^{\pm\frac{1}{2}}\xi^{\pm\frac{1}{2}}\label{bc06b}\\
\delta_{\lambda}\mathcal{Q}^{\pm\frac{1}{2}}	&=&	 \pm\frac{k}{\pi}\partial_{\varphi}\xi^{\pm\frac{1}{2}}-2\mathcal{L}^{\pm}\xi^{\mp\frac{1}{2}}
-2\mathcal{L}^{0}\xi^{\pm\frac{1}{2}}\pm\epsilon^{\pm}\mathcal{Q}^{\mp\frac{1}{2}}\pm\frac{1}{2}
\epsilon^{0}\mathcal{Q}^{\pm\frac{1}{2}}\label{bc06c}.
\end{eqnarray}
Analogously, the chemical potentials obey the following transformations
\begin{eqnarray}
\delta_\lambda \mu^{0}	&=&	 \partial_{t}\epsilon^{0}+2\mu^{+}\epsilon^{-}-2\mu^{-}\epsilon^{+}-2\nu^{\frac{1}{2}}\xi^{-\frac{1}{2}}-2\nu^{-\frac{1}{2}}\xi^{\frac{1}{2}}\label{bc07a}\\
\delta_\lambda \mu^{\pm}	&=&	 \partial_{t}\epsilon^{\pm}\mp\mu^{0}\epsilon^{\pm}\pm\mu^{\pm}\epsilon^{0}-2\nu^{\pm\frac{1}{2}}\xi^{\pm\frac{1}{2}}\label{bc07b}\\
\delta_\lambda \nu^{\pm\frac{1}{2}}	&=&	 \partial_{t}\xi^{\pm\frac{1}{2}}\mp\frac{1}{2}\mu^{0}\xi^{\pm\frac{1}{2}}\pm\mu^{\pm}\xi^{\mp\frac{1}{2}}\pm\frac{1}{2}\nu^{\pm\frac{1}{2}}\epsilon^{0}\mp\nu^{\mp\frac{1}{2}}\epsilon^{\pm}\label{bc07c}.
\end{eqnarray}
The left hand of these equations are all zero, in accordance with our boundary conditions.

The variation of the canonical boundary charge \eqref{sugra09} for the $\Gamma$-sector has the form
\begin{equation}\label{bc08}
\delta\mathbb{Q}=\int\mathrm{d}\varphi\;\left[\delta\mathcal{L}^{0}\epsilon^{0}
+\delta\mathcal{L}^{+}\epsilon^{-}+\delta\mathcal{L}^{-}\epsilon^{+}+\delta\mathcal{Q}^{\frac{1}{2}}
\xi^{-\frac{1}{2}}+\delta\mathcal{Q}^{-\frac{1}{2}}\xi^{\frac{1}{2}}\right].
\end{equation}
For the loosest set of boundary conditions, all parameters are considered as independents of the state-dependent functions $\left(\mathcal L^I,\mathcal Q^\alpha\right)$, in consequence, we can integrate the charge:
\begin{equation}\label{bc09}
\mathbb{Q}=\int\mathrm{d}\varphi\;\left[\mathcal{L}^{0}\epsilon^{0}+\mathcal{L}^{+}\epsilon^{-}+
\mathcal{L}^{-}\epsilon^{+}+\mathcal{Q}^{\frac{1}{2}}\xi^{-\frac{1}{2}}+\mathcal{Q}^{-\frac{1}{2}}
\xi^{\frac{1}{2}}\right]\end{equation}
The equation above states that all functions $\mathcal L^I$ and $\mathcal Q^\alpha$ are generators of the asymptotic symmetries. That is the reason why they are also called ``charges".

The asymptotic symmetry algebra can be obtained through the relation: $\delta_\lambda F = \left\{F,\mathbb{Q}\left[\lambda\right]\right\}_{\mathrm{P.B}}$, where for the left entry of this equation we use the results from \eqref{bc06a}, \eqref{bc06b} and \eqref{bc06c}. We obtain the following Poisson Brackets (P.B):
\begin{eqnarray}
\left\{ \mathcal{L}^{I}\left(t,\varphi\right),\mathcal{L}^{J}\left(t,\bar{\varphi}\right)\right\}_{\mathrm{P.B}}
&=& \left(I-J\right)\mathcal{L}^{I+J}\left(t,\varphi\right)\delta\left(\bar{\varphi}-\varphi\right)
+\frac{k}{2\pi}\kappa^{IJ}\partial_{\varphi}\delta\left(\bar{\varphi}-\varphi\right)\label{bc10a}\\
\left\{ \mathcal{L}^{I}\left(t,\varphi\right),\mathcal{Q}^{\alpha}\left(t,\bar{\varphi}\right)\right\}_{\mathrm{P.B}}	 &=&	 \left(\frac{I}{2}-\alpha\right)\mathcal{Q}^{I+\alpha}\left(t,\varphi\right)\delta\left(\bar{\varphi}-\varphi\right)\label{bc10b}\\
\left\{ \mathcal{Q}^{\alpha}\left(t,\varphi\right),\mathcal{Q}^{\beta}\left(t,\bar{\varphi}\right)\right\} _{\mathrm{P.B}}	&=&	 -2\mathcal{L}^{\alpha+\beta}\left(t,\varphi\right)\delta\left(\bar{\varphi}-\varphi\right)+\frac{k}{2\pi}
\kappa^{\alpha\beta}\partial_{\varphi}\delta\left(\bar{\varphi}-\varphi\right).\label{bc10c}
\end{eqnarray}
The supertrace between two bosonic or two fermionic elements of the $\mathfrak{osp}\left(1\mid 2\right)$ algebra are represented by $\kappa^{IJ}$ and $\kappa^{\alpha\beta}$, respectively. As shown in the appendix, there is no $\kappa^{I\alpha}$ element. Therefore, the presence of the $k$ term in  \eqref{bc10a} and \eqref{bc10c}, as well as its absence in  \eqref{bc10b}, hints the existence of a central extension. This can be shown explicitly if we decompose the charges in term of Fourier modes and use the correspondence principle to substitute the Poisson Brackets by anti-commutators, when both operators are fermionic, or by commutators otherwise. We obtain
\begin{eqnarray}
\left[\mathrm{L}_{n}^{I},\mathrm{L}_{m}^{J}\right] &=& \left(I-J\right)\mathrm{L}_{n+m}^{I+J}-kn\kappa^{IJ}\delta_{n+m,0}\label{bc11a}\\
\left[\mathrm{L}_{n}^{I},\mathrm{Q}_{r}^{\alpha}\right] &=& \left(\frac{I}{2}-\alpha\right)\mathrm{Q}_{n+r}^{I+\alpha}\label{bc11b}\\
\left\{\mathrm{Q}_{r}^{\alpha},\mathrm{Q}_{s}^{\beta}\right\} &=& -2\mathrm{L}_{r+s}^{\alpha+\beta}-kr\kappa^{\alpha\beta}\delta_{r+s,0}.\label{bc11c}
\end{eqnarray}
This is the $\mathfrak{osp}\left(1\mid 2\right)_k$ algebra. The indices of the bosonic fields $n,m$ are integers, while the indices of the fermionic fields $r,s$ can be integers or half-integers, related to Ramond (periodic) \cite{Ramond} or Neveu-Schwarz (anti-periodic) \cite{Neveu} sectors. The same algebra can also be obtained for the $\bar\Gamma$-sector. Therefore, the asymptotic symmetry algebra for the loosest set of boundary conditions of $\mathcal N=\left(1,1\right)$ supergravity is two copies of the $\mathfrak{osp}\left(1\mid 2\right)_k$ algebra.

To finish our discussion, let us talk about the conservation of the boundary charge. On-shell, we have that the time evolution of the boundary charge is given by:
\begin{equation}\label{bc12}
\partial_{t}\mathbb{Q}	=	 \frac{k}{2\pi}\int\mathrm{d}\varphi\;\left[\frac{1}{2}\partial_{\varphi}\mu^{0}\epsilon^{0}
-\partial_{\varphi}\mu^{+}\epsilon^{-}-\partial_{\varphi}\mu^{-}\epsilon^{+}+2\partial_{\varphi}
\nu^{\frac{1}{2}}\xi^{-\frac{1}{2}}-2\partial_{\varphi}\nu^{-\frac{1}{2}}\xi^{\frac{1}{2}}\right].
\end{equation}
When the chemical potentials are constants or $\varphi$-independents, the boundary charge is conserved: $\partial_{t}\mathbb{Q}=0$. On the other hand, if the chemical potentials depend on the angular variable $\varphi$, we have $\partial_t \delta \mathbb{Q} = 0$, once that the chemical potentials are not allowed to vary. Therefore, in both cases, we have a conservation law for the boundary charge.

\section{Other boundary conditions for supergravity:}\label{sec4}
In this section, we will impose certain conditions on the gauge connection components: \eqref{bc01} and \eqref{bc02} that lead us to interesting asymptotic symmetry algebras.
We will show three cases: For the supersymmetric version of Brown-Henneaux boundary conditions we obtain the super-Virasoro algebra (also called $N=1$ super-conformal algebra), for the super-symmetric version of Troessaert boundary conditions we obtain the super-Virasoro algebra plus an $\mathfrak{u}(1)$ current algebra, and for the supersymmetric version of the Avery-Poojary-Suryanarayana boundary conditions, the semidirect sum of a $\mathfrak{sl}\left(2\right)_k$ and a super-Virasoro algebra.

\subsection{Super-conformal boundary conditions:}

Super-conformal boundary conditions are the supersymmetric extension of the usual Brown-Henneaux boundary conditions for $\mathrm{AdS_3}$ gravity. In this case, the angular component of the auxiliary gauge connection takes the form
\begin{equation}
a_\varphi=L_1+S,\;\;\;\;\mathrm{where}\;\;\left[L_{-1},S\right]=0. \label{hw}
\end{equation}
This condition is the so called highest weight ansatz and it set the fields $\mathcal{L}^{0}$, $\mathcal{Q}^{\frac{1}{2}}$ to zero and $\mathcal{L}^{+}=-\frac{k}{2\pi}$. The only dynamical fields are $\mathcal L \equiv \mathcal L^-$ and $\mathcal Q \equiv \mathcal Q^{-\frac{1}{2}}$. Under this condition we have: 
\begin{eqnarray}
a_\varphi &=& L_1 - \frac{2\pi}{k}\mathcal L L_{-1} - \frac{\pi}{k}\mathcal Q G_{-\frac{1}{2}}\label{bc13a}\\
a_t &=& \mu L_1 - \mu' L_0 + \left(\frac{1}{2}\mu''-\frac{2\pi}{k}\mathcal L \mu - \frac{\pi}{k}\mathcal Q \nu\right)L_{-1} + \nu G_{\frac{1}{2}} - \left( \nu' + \frac{\pi}{k}\mu\mathcal Q\right) G_{-\frac{1}{2}}\label{bc13b}
\end{eqnarray}
where $\mu\equiv\mu^{+}$, $\nu\equiv\nu^{\frac{1}{2}}$ are the independent chemical potentials. The primes denote the angular derivatives. Furthermore, from the equation of motion we obtain
\begin{eqnarray}
\partial_{t}\mathcal{L}	&=&	 -\frac{k}{4\pi}\mu'''+2\mathcal{L}\mu'+\mathcal{L}'\mu+\frac{3}{2}Q\nu'+\frac{1}{2}Q'\nu\label{bc14a}\\
\partial_{t}Q &=& \frac{k}{\pi}\nu''-2\mathcal{L}\nu+\frac{3}{2}\mu'Q+\mu Q'.\label{bc14b}
\end{eqnarray}
Following a similar approach as Brown and Henneaux, we choose $\mu=1$ and $\nu=0$. Therefore, the highest weight ansatz and an appropriated choice of chemical potentials  lead us to $a_\varphi=a_t$, as in the pure bosonic case. Moreover, for this choice of chemical potentials the equations of motion are reduced to $\partial_\tau \mathcal L = \mathcal L'$ and $\partial_\tau \mathcal Q = \mathcal Q'$.

The conditions $\delta_\lambda \mathcal L^0=\delta_\lambda \mathcal Q^{\frac{1}{2}}=\delta_\lambda \mathcal L^+=0$ imply that the only independent parameters are $\epsilon\equiv\epsilon^+$ and $\xi=\xi^{\frac{1}{2}}$. In fact, the gauge parameter has the form:
\begin{eqnarray}\label{bc15}
\lambda = b^{-1}\left[\epsilon L_1 - \epsilon'L_0 + \left(\frac{1}{2}\epsilon''-\frac{2\pi}{k}\mathcal{L}\epsilon-\frac{\pi}{k}Q\xi\right)L_{-1} + \xi G_{\frac{1}{2}} - \left( \xi'+\frac{\pi}{k}\epsilon Q\right) G_{-\frac{1}{2}}\right]b.
\end{eqnarray}
Replacing this in the transformation of the fields $\mathcal L$ and $\mathcal Q$, we obtain
\begin{eqnarray}
\delta\mathcal{L} &=&	 -\frac{k}{4\pi}\epsilon'''+2\mathcal{L}\epsilon'+\mathcal{L}'\epsilon+\frac{3}{2}Q\xi'+\frac{1}{2}Q'\xi\label{bc16a}\\
\delta\mathcal{Q} &=& \frac{k}{\pi}\xi''-2\mathcal{L}\xi+\frac{3}{2}\epsilon'Q+\epsilon Q'.\label{bc16b}
\end{eqnarray}
These transformations hint the super-Virasoro algebra \cite{Banados-SUSY}. Now, we proceed to derive the asymptotic symmetry algebra. First, we notice that the variation of the boundary charge is reduced to  $\delta\mathbb{Q}=\int\mathrm{d}\varphi\;\left[\delta\mathcal{L}\epsilon+\delta\mathcal{Q}\xi\right]$ and can be integrated since $\epsilon$ and $\xi$ are state independent:
\begin{equation}
\mathbb{Q} = \int\mathrm{d}\varphi\;\left[\mathcal{L}\epsilon+\mathcal{Q}\xi\right].\label{bc17}
\end{equation}
We identify $\mathcal{L}$ and $\mathcal{Q}$ as generators of the asymptotic symmetries. Then, we decompose in Fourier modes, we obtain:
\begin{eqnarray}
\left[\mathrm{L}_{n},\mathrm{L}_{m}\right] &=& \left(n-m\right)\mathrm{L}_{n+m}+\frac{k}{2}n^{3}\delta_{n+m,0}\label{bc18a}\\
\left[\mathrm{L}_{n},\mathrm{Q}_{r}\right] &=& \left(\frac{n}{2}-r\right)\mathrm{Q}_{n+r}\label{bc18b}\\
\left\{\mathrm{Q}_{r},\mathrm{Q}_{s}\right\} &=& 2\mathrm{L}_{r+s}+2kr^{2}\delta_{r+s,0}.\label{bc18c}
\end{eqnarray}
This is the super-Virasoro algebra with central charge $c=6k$. The usual factors in \eqref{bc18a} and \eqref{bc18c} are obtained by shifting the zero mode $L_0$.

For the $\bar\Gamma$-sector, we choose $\bar{\mathcal{L}}^{0}=0=\bar{\mathcal{Q}}^{-\frac{1}{2}}$ and $\bar{\mathcal{L}}^{-}=\frac{k}{2\pi}$. These conditions also lead to a super-Virasoro algebra. Therefore, the asymptotic symmetry algebra of the supersymmetric extension of the Brown-Henneaux boundary conditions is two copies of the super-Virasoro algebra.

\subsection{Warped super-conformal boundary conditions:}

The Troessaert boundary conditions are a generalization of the Brown-Henneaux boundary conditions. They allow a conformal factor in the leading order of the metric \eqref{sugra08} and lead to an enhancement of the asymptotic symmetry algebra \cite{Troessaert}, generated by two Virasoro algebras and two $\mathfrak{u}\left(1\right)$ algebras. Detournay and Riegler \cite{Detournay} also showed new boundary conditions that are equivalent to the Troessaert ones in the flat space limit: When the cosmological constant goes to zero, the asymptotic symmetry algebra consists in two copies of the $\mathrm{bms_3}$ algebra with two $\mathfrak{u}\left(1\right)$ algebras.

In this case, let us work with a supersymmetric extension of the Troessaert boundary conditions. First, let us set all chemical potentials equal to zero $\mu^I=\nu^\alpha=0$ and for the angular components of the auxiliary connection we goes beyond the highest weight ansatz \eqref{hw}, allowing the field $\mathcal L^+$ to fluctuate. This is, we fix $\mathcal{L}^{0}=\mathcal{Q}^{\frac{1}{2}}=0$ and leave $\mathcal L^+,\mathcal L^-$ and $\mathcal Q^{-\frac{1}{2}}$ to vary. However, we choose a particular expression for $\mathcal L^+$: $\mathcal{L}^{+}=-\frac{k}{2\pi}e^{2\phi}$ where $\phi=\phi\left(\varphi\right)$. In fact, the function $e^{2\phi}$ is the conformal factor responsible for the fluctuation of the leading order of the metric. It is important to define the following quantities
\begin{equation}\label{bc19}
\mathcal{L}\equiv e^{2\phi}\mathcal{L}^{-},\;\;\;\mathcal{J}\equiv-\frac{k}{2\pi}\phi',\;\;\;\mathcal{Q}\equiv e^{\phi}\mathcal{Q}^{-\frac{1}{2}}.
\end{equation}
The auxiliary gauge connection $a_\varphi$ has the form
\begin{equation}\label{bc20}
a_{\varphi}\left(t,\varphi\right)=e^{2\phi}L_{1}-\frac{2\pi}{k}e^{-2\phi}\mathcal{L}L_{-1}-\frac{\pi}{k}e^{-\phi}\mathcal{Q}G_{-\frac{1}{2}}.
\end{equation}
Note that, for $\phi=0$, the connection $a_\varphi$ coincides with the one given in \eqref{bc13a}. However, the conditions for the chemical potential are different from the ones in the previous subsection.

As a result of imposing $\delta\mathcal L^0 =\delta\mathcal Q^{\frac{1}{2}}=0$, the gauge parameter $\lambda$ only has three independent components, denoted by $\epsilon,\epsilon^0$ and $\xi$:
\begin{eqnarray}\label{bc21}
\lambda &=& b^{-1}\left[e^{2\phi} \epsilon L_1 + \epsilon^0 L_0 	 -e^{-2\phi}\left(\frac{1}{2}\epsilon^{0}{}'+\frac{2\pi}{k}\mathcal{L}\epsilon+\frac{\pi}{k}Q\xi\right)L_{-1}\right. \nonumber\\ &&  \left. + e^\phi\xi G_{\frac{1}{2}} -  e^{-\phi}\left(\xi'-\frac{2\pi}{k}\mathcal{J}\xi+\frac{\pi}{k}\epsilon Q\right)G_{-\frac{1}{2}}\right]b.
\end{eqnarray}

Therefore, the transformations of the quantities \eqref{bc19} are
\begin{eqnarray}
\delta\mathcal{J} &=& -\frac{k}{4\pi}\epsilon^{0}{}'-\frac{k}{4\pi}\epsilon''+\epsilon'\mathcal{J}+\epsilon \mathcal{J}'\label{bc22a}\\
\delta\mathcal{L} &=& \frac{k}{4\pi}\epsilon^{0}{}''+\mathcal{J}\epsilon^{0}{}'+2\mathcal{L}\epsilon'+\mathcal{L}'\epsilon+
\frac{3}{2}Q\xi'+\frac{1}{2}Q'\xi\label{bc22b}\\
\delta\mathcal{Q} &=&\frac{k}{\pi}\xi''-2\mathcal{N} \xi+\frac{3}{2}\epsilon' Q+\epsilon Q'\label{bc22c}
\end{eqnarray}
where
\begin{equation}\label{bc24}
\mathcal{N}\equiv\mathcal{L}+\mathcal{J}'+\frac{2\pi}{k}\mathcal{J}^{2}.
\end{equation}
Note that $\mathcal J$ is unaffected by the supersymmetry, once that its transformation is identical to the pure bosonic case \cite{Grumiller3D}. The best way to read off this algebra is to go to Fourier modes, but first we need the boundary charge. Since
\begin{equation}\label{bc25}
\delta\mathbb{Q} = \int\mathrm{d}\varphi\;\left[\delta\mathcal{J}\epsilon^{0}
+\delta\mathcal{L}\epsilon+\delta\mathcal{Q}\xi\right]
\end{equation}
and the parameters $\epsilon^0,\epsilon$ and $\xi$ are field independent, the above expression can be functionally integrated. We identify $\mathcal{L}$ , $\mathcal{J}$ and $\mathcal{Q}$ as the generators of the asymptotic symmetries.
Using the Fourier mode decomposition, we obtain
\begin{eqnarray}
\left[\mathrm{L}_{n},\mathrm{L}_{m}\right]&=&\left(n-m\right)\mathrm{L}_{n+m}\label{bc26a}\\
\left[\mathrm{L}_{n},\mathrm{J}_{m}\right]&=&-m\mathrm{J}_{n+m}+\frac{k}{2}in^{2}\delta_{n+m,0}\label{bc26b}\\
\left[\mathrm{J}_{n},\mathrm{J}_{m}\right]&=&-\frac{k}{2}n\delta_{m+n,0}\label{bc26c}\\
\left[\mathrm{L}_{n},\mathrm{Q}_{r}\right]&=&\left(\frac{n}{2}-r\right)\mathrm{Q}_{n+r}\label{bc26d}\\
\left\{\mathrm{Q}_{r},\mathrm{Q}_{s}\right\}&=&2\mathrm{N}_{r+s}+2kr^{2}\delta_{r+s,0}.\label{bc26e}
\end{eqnarray}
This is the supersymmetric extension of the algebra presented by Troessaert in \cite{Troessaert}. Note that, in this case,  $\mathrm{L}_n$ satisfy a Virasoro algebra without central extensions \eqref{bc26a}, $\mathrm{J}_{n}$ are the $\mathfrak{u}_1$ generators and present an anomalous term in \eqref{bc26b}. The brackets \eqref{bc26c} determine the Kac-Moody level, which in this case is $\kappa=-k$. Also notice that the presence of $\mathrm{N}_n$ (the Fourier modes of $\mathcal N$) instead of $\mathrm{L}_n$ in the anti-commutator of the supercharge $\mathrm{Q}_r$, this is consequence of the transformation of $\mathcal N$
\begin{eqnarray}
\delta \mathcal N = -\frac{k}{4\pi}\epsilon'''+2\mathcal{N}\epsilon'+\mathcal{N}'\epsilon+\frac{3}{2}Q\xi'+\frac{1}{2}Q'\xi\label{bc27}
\end{eqnarray}
which, along with $\mathcal Q$, transforms as a super-Virasoro with central extension.

This warped super-conformal algebra is also obtained for the $\bar\Gamma$ sector if we choose $\bar{\mathcal L}^0=0$, $\bar{\mathcal L}^- = \frac{k}{2\pi}e^{2\bar{\phi}}$ and $\bar{\mathcal{Q}}^{-\frac{1}{2}}=0$. Therefore, the asymptotic symmetry algebra of $\mathcal N=\left(1,1\right)$ supergravity with Troessaert boundary conditions is two copies of the semidirect sum of super-Virasoro algebra and an $\mathfrak{u}(1)$ algebra.

\subsection{Avery-Poojary-Suryanarayana boundary conditions for $\mathcal N=1$ supergravity}

For pure Chern-Simons gravity, the Avery-Poojary-Suryanarayana boundary conditions \cite{Avery} can be obtained by choosing, for the $\Gamma$-sector, the Brown-Henneaux boundary conditions and leaving all charges to vary for the $\bar{\Gamma}$-sector. As a result, the asymptotic symmetry is the semidirect sum of a Virasoro algebra and a $\mathfrak{sl}(2)_k$ algebra.

Now, for $\mathcal N=1$ supergravity, or $\mathcal N=\left(1,0\right)$ supergravity, we need to work with the $\mathfrak{osp}\left(1\mid2\right)\otimes\mathfrak{sl}\left(2\right)$ algebra, this means that our previous formulas are valid if we set all fermions fields to zero in the $\bar{\Gamma}$-sector, where the dynamical fields are written in $\mathfrak{sl}\left(2\right)$ basis. Therefore, the supersymmetric version of the Avery-Poojary-Suryanarayana boundary conditions for $\mathcal N=1$ supergravity is given by
\begin{eqnarray}
\Gamma-\mathrm{sector}&:& \mathcal L^0=\mathcal Q^{\frac{1}{2}}=0,\;\; \mathcal L^+=-\frac{k}{2\pi}\;\; \mathrm{and}\;\; \mathcal L^-,\;\mathcal Q\equiv \mathcal Q^{-\frac{1}{2}}\;\; \mathrm{vary}\label{bc28a}\\
\bar{\Gamma}-\mathrm{sector}&:& \bar{\mathcal Q}^{\frac{1}{2}}=\bar{\mathcal Q}^{-\frac{1}{2}}=0,\;\mathrm{and\;all}\;\; \mathcal{T}^I\equiv\bar{\mathcal{L}}^I\;\; \mathrm{vary}\label{bc28b}
\end{eqnarray}
this is, supersymmetric Brown-Henneaux boundary for the $\Gamma$-sector and the loosest set of boundary conditions for the $\bar\Gamma$-sector. Then, we perform the Sugawara shift (as shown in \cite{Grumiller3D} for $\mathrm{AdS}_3$ gravity):
\begin{equation}
    \mathcal{L}\equiv\mathcal{L}^{-}+\frac{2\pi}{k}\left(\mathcal{T}^{0}\mathcal{T}^{0}-\mathcal{T}^{+}\mathcal{T}^{-}\right).\label{bc29}
\end{equation}
Therefore, we obtain the following algebra:
\begin{eqnarray}
\left[ \mathrm{L}_{n},\mathrm{L}_{m}\right] &=&\left(n-m\right)\mathrm{L}_{n+m}+\frac{k}{2}n^{3}\delta_{n+m,0}\label{bc30a}\\
\left[\mathrm{L}_{n},\mathrm{Q}_{r}\right]&=&\left(\frac{n}{2}-r\right)\mathrm{Q}_{n+r}\label{bc30b}\\
\left\{\mathrm{Q}_{r},\mathrm{Q}_{s}\right\} &=& 2\mathrm{L}_{r+s}+2kr^{2}\delta_{r+s,0}\label{bc30c}\\
\left[\mathrm{L}_{n},\mathrm{T}_{m}^{I}\right]&=&-m\mathrm{T}_{n+m}^{I}\label{bc30d}\\
\left[\mathrm{T}_{n}^{I},\mathrm{T}_{m}^{J}\right]&=&\left(I-J\right)\mathrm{T}_{n+m}^{I+J}-nk\kappa^{IJ}\delta_{n+m,0}.\label{bc30e}
\end{eqnarray}
This is the semidirect sum of a super-Virasoro algebra and the $\mathrm{sl}(2)_k$ algebra.
The supersymmetric extensions of the Avery-Poojary-Suryanarayana boundary conditions for $\mathcal N=\left(1,1\right)$ supergravity were studied by Poojary-Suryanarayana in \cite{Poojary}.

\section{Most general boundary condition  for $\mathcal N= \left(2,2\right)$ supergravity}\label{sec5}

In \cite{Gunaydin} was shown a classification of consistent extended supergravity theories. However, for some models, their asymptotic symmetries are related super-conformal algebras with non-linearities. To avoid those non-linearities, we will work with $\mathfrak{osp}\left(2\mid2\right)\otimes\mathfrak{osp}\left(2\mid2\right)$ Chern-Simons supergravity. However, since the super-algebra $\mathfrak{osp}\left(2\mid2\right)$ contains the $\mathfrak{sl}\left(2\right)$ algebra, this section is valid for $\mathcal N=2$ supergravity as well.

Let us begin writing the auxiliary gauge connection in terms of the generators of the $\mathfrak{osp}\left(2\mid 2\right)$ algebra:
\begin{eqnarray}
a_{\varphi}\left(t,\varphi\right) &=&	 -\frac{2\pi}{k}\left[\mathcal{L}^{+}L_{1}-2\mathcal{L}^{0}L_{0}+\mathcal{L}^{-}L_{-1}-\frac{1}{2}\mathcal{Q}_{A}^{\frac{1}{2}}G_{\frac{1}{2}}^{A}+\frac{1}{2}\mathcal{Q}_{A}^{-\frac{1}{2}}G_{-\frac{1}{2}}^{A}-\frac{1}{2}\mathcal{B}T\right]\label{ebc01a}\\
a_{t}\left(t,\varphi\right) &=& \mu^{I}L_{I}+\nu_{A}^{\alpha}G_{\alpha}^{A}+\sigma T.\label{ebc01b}
\end{eqnarray}
Note that, in \eqref{ebc01a}, we have three $\mathfrak{sl}(2)$ fields $\mathcal L^I$, four fermionic fields $\mathcal Q^\alpha_A$ ($A=1,2$), and one field $\mathcal B$ related to the $\mathfrak{so}(2)$ algebra. Analogously, in \eqref{ebc01b} we have introduced eight chemical potentials $\left(\mu^I,\nu^\alpha_A,\sigma\right)$. Remember that for the $\bar\Gamma$-sector, we just need to change $k$ by $-k$ and change all fields and chemical potential by barred ones. Therefore, for $\mathcal{N}=\left(2,2\right)$ gravity we need to introduce sixteen varying functions $\left(\mathcal{L}^I,Q^\alpha_A,\mathcal B,\bar{\mathcal L}^I,\bar{Q}^\alpha_A,\bar{\mathcal B}\right)$, and sixteen chemical potential.

For the loosest set of boundary conditions, the equation of motion becomes
\begin{eqnarray}
\partial_{t}\mathcal{L}^{0}	&=&	 \frac{k}{4\pi}\partial_{\varphi}\mu^{0}-\mathcal{L}^{+}\mu^{-}+\mathcal{L}^{-}\mu^{+}-\frac{1}{2}\mathcal{Q}_{A}^{\frac{1}{2}}\nu_{A}^{-\frac{1}{2}}+\frac{1}{2}\mathcal{Q}_{A}^{-\frac{1}{2}}\nu_{A}^{\frac{1}{2}}\label{ebc02a}\\
\partial_{t}\mathcal{L}^{\pm}	&=&	 -\frac{k}{2\pi}\partial_{\varphi}\mu^{\pm}\pm2\mathcal{L}^{0}\mu^{\pm}\pm\mathcal{L}^{\pm}\mu^{0}\pm\mathcal{Q}_{A}^{\pm\frac{1}{2}}\nu_{A}^{\pm\frac{1}{2}}\label{ebc02b}\\
\partial_{t}\mathcal{Q}_{A}^{\pm\frac{1}{2}}	&=&	 \pm\frac{k}{\pi}\partial_{\varphi}\nu_{A}^{\frac{1}{2}}-2\mathcal{L}^{\pm}\nu_{A}^{\mp\frac{1}{2}}-2\mathcal{L}^{0}\nu_{A}^{\pm\frac{1}{2}}\pm\mathcal{Q}_{A}^{\mp\frac{1}{2}}\mu^{+}\pm\frac{1}{2}\mu^{0}\mathcal{Q}_{A}^{\pm\frac{1}{2}}\nonumber\\
&&\pm\epsilon_{AB}\mathcal{B}\nu_{B}^{\pm\frac{1}{2}}-\epsilon_{AB}\sigma\mathcal{Q}_{B}^{\pm\frac{1}{2}}\label{ebc02c}\\
\partial_{t}\mathcal{B}	&=&	 \frac{k}{\pi}\partial_{\varphi}\sigma-\epsilon_{AB}\mathcal{Q}_{A}^{\frac{1}{2}}\nu_{B}^{-\frac{1}{2}}-\epsilon_{AB}\mathcal{Q}_{A}^{-\frac{1}{2}}\nu_{B}^{\frac{1}{2}}.\label{ebc02d}
\end{eqnarray}
Note that the equation of motion for the gauge fields $\mathcal L^I$ are unaffected by the chemical potential $\sigma$, related to the $\mathrm{SO}(2)$ group, explicitly.

The gauge parameter can be written as
\begin{equation}
\lambda=b^{-1}\left[\epsilon^{I}L_{I}+\xi^{\alpha}_A G^A_{\alpha}+\chi T\right]b\label{ebc03}
\end{equation}
where $\epsilon^I,\;\xi^\alpha_A$ and $\chi$ are considered field independents. This consideration results useful when we integrate the boundary charge.

The gauge transformation preserving the form of the boundary condition \eqref{ebc01a} are
\begin{eqnarray}
\delta_{\lambda}\mathcal{L}^{0}	&=&	 \frac{k}{4\pi}\partial_{\varphi}\epsilon^{0}-\mathcal{L}^{+}\epsilon^{-}+\mathcal{L}^{-}\epsilon^{+}-\frac{1}{2}\mathcal{Q}_{A}^{\frac{1}{2}}\xi_{A}^{-\frac{1}{2}}+\frac{1}{2}\mathcal{Q}_{A}^{-\frac{1}{2}}\xi_{A}^{\frac{1}{2}}\label{ebc04a}\\
\delta_{\lambda}\mathcal{L}^{\pm}	&=&	 -\frac{k}{2\pi}\partial_{\varphi}\epsilon^{\pm}\pm2\mathcal{L}^{0}\epsilon^{\pm}\pm\mathcal{L}^{\pm}\epsilon^{0}\pm\mathcal{Q}_{A}^{\pm\frac{1}{2}}\xi_{A}^{\pm\frac{1}{2}}\label{ebc04b}\\
\delta_{\lambda}\mathcal{Q}_{A}^{\pm\frac{1}{2}}	&=&	 \pm\frac{k}{\pi}\partial_{\varphi}\xi_{A}^{\pm\frac{1}{2}}-2\mathcal{L}^{\pm}\xi_{A}^{\mp\frac{1}{2}}-2\mathcal{L}^{0}\xi_{A}^{\pm\frac{1}{2}}\pm\mathcal{Q}_{A}^{\mp\frac{1}{2}}\epsilon^{\pm}\pm\frac{1}{2}\mathcal{Q}_{A}^{\pm\frac{1}{2}}\epsilon^{0}\nonumber\\
&&\pm\epsilon_{AB}\mathcal{B}\xi_{B}^{\pm\frac{1}{2}}-\epsilon_{AB}\mathcal{Q}_{B}^{\pm\frac{1}{2}}\chi\label{ebc04c}\\
\delta_{\lambda}\mathcal{B}	&=&	 \frac{k}{\pi}\partial_{\varphi}\chi-\epsilon_{AB}\mathcal{Q}_{A}^{\frac{1}{2}}\xi_{B}^{-\frac{1}{2}}-\epsilon_{AB}\mathcal{Q}_{A}^{-\frac{1}{2}}\xi_{B}^{\frac{1}{2}}\label{ebc04d}
\end{eqnarray}
analogously, for the chemical potentials \eqref{ebc01b} we have
\begin{eqnarray}
\delta_{\lambda}\mu^{0}&=&	 \partial_t\epsilon^{0}+2\mathcal{\mu}^{+}\epsilon^{-}-2\mu^{-}\epsilon^{+}-2\nu_{A}^{\frac{1}{2}}\xi_{A}^{-\frac{1}{2}}-2\nu_{A}^{-\frac{1}{2}}\xi_{A}^{\frac{1}{2}}\label{ebc05a}\\
\delta_{\lambda}\mu^{\pm}&=&	 \partial_t\epsilon^{\pm}\mp\mu^{0}\epsilon^{\pm}\pm\mu^{\pm}\epsilon^{0}-2\nu_{A}^{\pm\frac{1}{2}}\xi_{A}^{\pm\frac{1}{2}}\label{ebc05b}\\
\delta_{\lambda}\nu_{A}^{\pm\frac{1}{2}}&=&	 \partial_t\xi_{A}^{\pm\frac{1}{2}}\pm\mu^{\pm}\xi_{A}^{\mp\frac{1}{2}}\mp\frac{1}{2}\mu^{0}\xi_{A}^{\pm\frac{1}{2}}\mp\epsilon^{\pm}\nu_{A}^{\mp\frac{1}{2}}\pm\frac{1}{2}\epsilon^{0}\nu_{A}^{\pm\frac{1}{2}}+\epsilon_{AB}\sigma\xi_{B}^{\pm\frac{1}{2}}-\epsilon_{AB}\chi\nu_{B}^{\pm\frac{1}{2}}\label{ebc05c}\\
\delta_{\lambda}\sigma&=&	 \partial_t\chi-\epsilon_{AB}\nu_{A}^{\frac{1}{2}}\xi_{B}^{-\frac{1}{2}}+\epsilon_{AB}\nu_{A}^{-\frac{1}{2}}\xi_{B}^{\frac{1}{2}}\label{ebc05d}.
\end{eqnarray}
Remember that the left side of equations \eqref{ebc05a}-\eqref{ebc05d} are zero, once that chemical potential are not allowed to vary.

The canonical boundary charge is now given by
\begin{eqnarray}
\delta_\lambda \mathbb{Q}=\int\mathrm{d}\varphi\;\left[\delta_\lambda\mathcal{L}^{0}\epsilon^{0}+\delta_\lambda\mathcal{L}^{+}\epsilon^{-}+\delta_\lambda\mathcal{L}^{-}\epsilon^{+}+\delta_\lambda\mathcal{Q}^{\frac{1}{2}}_A\xi^{-\frac{1}{2}}_A+\delta_\lambda\mathcal{Q}^{-\frac{1}{2}}_A\xi^{\frac{1}{2}}_A+\delta_\lambda\mathcal{B}\chi\right]\label{ebc06}
\end{eqnarray}
which, for the loosest set of boundary conditions, can be integrated. After the integration we identify all fields $\left(\mathcal L^I, \mathcal Q^\alpha_A,\mathcal B\right)$ as charges. Performing the Fourier mode decomposition, we obtain
\begin{eqnarray}
 \left[\mathrm{L}_{n}^{I},\mathrm{L}_{m}^{J}\right]	&=&	 \left(I-J\right)\mathrm{L}_{n+m}^{I+J}-kn\kappa^{IJ}\delta_{n+m,0}\label{ebc07a}\\
\left[\mathrm{L}_{n}^{I},\mathrm{Q}_{Ar}^{\alpha}\right]	&=&	 \left(\frac{I}{2}-\alpha\right)\mathrm{Q}_{A\left(n+r\right)}^{I+\alpha}\label{ebc07b}\\
\left[\mathrm{Q}_{Ar}^{\alpha},\mathrm{B}_{m}\right]	&=&	 \epsilon_{AB}\mathrm{Q}_{B\left(r+m\right)}^{\alpha}\label{ebc07c}\\
\left[\mathrm{B}_{n},\mathrm{B}_{m}\right]	&=&	-kn\kappa \delta_{n+m,0}\label{ebc07d}\\
\left\{ \mathrm{Q}_{Ar}^{\alpha},\mathrm{Q}_{Bs}^{\beta}\right\} 	&=&	 -2\delta_{AB}\mathrm{L}_{r+s}^{\alpha+\beta}-\left(\alpha-\beta\right)\epsilon_{AB}\mathrm{B}_{r+s}-kr\kappa_{AB}^{\alpha\beta}\delta_{r+s,0}.\label{ebc07e}
\end{eqnarray}
This is the $\mathfrak{osp}\left(2\mid 2\right)_k$ super-algebra. As we can see, the central extension appears explicitly in \eqref{ebc07a}, \eqref{ebc07d} and the anti-commutator \eqref{ebc07e}.

The analysis of conservation of the canonical boundary charge follows the same procedure that in section 4: We need to derive with respect to time the boundary charge and replace the equations of motions. Then, when the chemical potentials are $\varphi$-independent, $\mathbb{Q}$ is conserved. Otherwise, $\delta_\lambda\mathbb{Q}$ is conserved.

\section{Other boundary conditions for extended supergravity}\label{sec6}

In this section we will present two boundary conditions for $\mathcal N=\left(2,2\right)$ supergravity: The extended supersymmetric version of Brown-Henneaux and Troessaert. We obtain, as asymptotic symmetry algebra, the $N=2$ super-conformal algebra, the $N=2$ warped super-conformal algebra. For $\mathcal N=2$ supergravity, the supersymmetric Avery-Poojary-Suryanarayana boundary conditions lead to a semidirect sum of a $N=2$ super-conformal algebra and a $\mathfrak{sl}\left(2\right)_k$ algebra as an asymptotic symmetry.

\subsection{$N=2$ super-conformal boundary conditions:}
In this case we impose the highest weight ansatz \eqref{hw}, which set $\mathcal L^0 = \mathcal Q^{\frac{1}{2}}_A=0$ and $\mathcal L^+=-\frac{k}{2\pi}$, while allows $\mathcal L \equiv \mathcal L^-$, $\mathcal Q_A \equiv \mathcal Q^{-\frac{1}{2}}_A$ and $\mathcal B$ to vary. In this case, the gauge parameter only have four independent components: $\epsilon\equiv\epsilon^+$, $\xi_A\equiv\xi^\frac{1}{2}_A$ and $\chi$. Explicitly, we have
\begin{eqnarray}\label{ebc08}
\lambda &=& b^{-1}\left[\epsilon L_{1}-\epsilon'L_{0}+\left(\frac{1}{2}\epsilon''-\frac{2\pi}{k}\mathcal{L}\epsilon-\frac{\pi}{k}\mathcal{Q}_{A}\xi_{A}\right)L_{-1}+\xi_{A}G_{\frac{1}{2}}^{A}\right.\nonumber\\
&&\left.-\left(\xi_{A}'+\frac{\pi}{k}\mathcal{Q}_{A}\epsilon+\frac{\pi}{k}\epsilon_{AB}\mathcal{B}\xi_{B}\right)G_{-\frac{1}{2}}^{A}+\chi T\right]b.
\end{eqnarray}
The transformations for $\left(\mathcal L, \mathcal Q, \mathcal B\right)$ are
\begin{eqnarray}
\delta_{\lambda}\mathcal{L}&=&-\frac{k}{4\pi}\epsilon'''
+2\mathcal{L}\epsilon'+\mathcal{L}'\epsilon
+\frac{3}{2}\mathcal{Q}_{A}\xi_{A}'+\frac{1}{2}\mathcal{Q}_{A}'\xi_{A}+\frac{\pi}{k}\epsilon_{AB}\mathcal{Q}_{A}\mathcal{B}
\xi_{B}\label{ebc09a}\\
\delta_{\lambda}\mathcal{Q}_{A}&=&\frac{k}{\pi}\xi_{A}''
-2\left(\mathcal{L}+\frac{\pi}{2k}\mathcal{B}^{2}\right)\xi_{A}+\frac{3}{2}\mathcal{Q}_{A}\epsilon'
+\mathcal{Q}_{A}'\epsilon+2\epsilon_{AB}\mathcal{B}\xi_{B}'
\nonumber\\
&&+\epsilon_{AB}\mathcal{B}'\xi_{B}+\epsilon_{AB}\mathcal{Q}_{B}\left(\frac{\pi}{k}\mathcal{B}\epsilon-\chi\right)\label{ebc09b}\\
\delta_{\lambda}\mathcal{B}&=&\frac{k}{\pi}\chi'-\epsilon_{AB}\mathcal{Q}_{A}\xi_{B}.\label{ebc09c}
\end{eqnarray}
For these boundary conditions, the boundary charge \eqref{ebc06} is integrable and we identify $\mathcal L,\;\mathcal Q$ and $\mathcal B$ as charges. However, these charges does not generate an algebra due to the last term on \eqref{ebc09a}, which is not linear. Moreover, the transformation $\mathcal L$ does not depend on the parameter $\chi$ related to the $\mathfrak{so}(2)$ fields. This means that the Fourier modes of $\mathcal L$ does not act on the Fourier modes of $\mathcal B$. To obtain the correct super-algebra, we need to perform a shift on $\mathcal L$ and a redefinition of the gauge parameter $\chi$:
\begin{eqnarray}
\mathcal L\rightarrow \hat{\mathcal L} \equiv \mathcal{L}+\frac{\pi}{2k}\mathcal{B}^{2},\;\;\;\;\;\;\eta\equiv \chi-\frac{\pi}{k}\mathcal{B}\epsilon.\label{ebc10}
\end{eqnarray}
Under these new variables, the gauge transformations \eqref{ebc09a}-\eqref{ebc09c} becomes:
\begin{eqnarray}
\delta_{\lambda}\hat{\mathcal{L}}&=&-\frac{k}{4\pi}\epsilon'''+2\hat{\mathcal{L}}\epsilon'+\hat{\mathcal{L}}'\epsilon
+\frac{3}{2}\mathcal{Q}_{A}\xi_{A}'+\frac{1}{2}\mathcal{Q}_{A}'\xi_{A}+\mathcal{B}\eta'\label{ebc11a}\\
\delta_{\lambda}\mathcal{Q}_{A}&=&\frac{k}{\pi}\xi_{A}''-2\hat{\mathcal{L}}\xi_{A}+\frac{3}{2}\mathcal{Q}_{A}\epsilon'
+\mathcal{Q}_{A}'\epsilon+2\epsilon_{AB}\mathcal{B}\xi_{B}'+\epsilon_{AB}\mathcal{B}'\xi_{B}-\epsilon_{AB}\mathcal{Q}_{B}\eta\label{ebc11b}\\
\delta_{\lambda}\mathcal{B}&=&\frac{k}{\pi}\eta'+\mathcal{B}\epsilon'+\mathcal{B}'\epsilon-\epsilon_{AB}
\mathcal{Q}_{A}\xi_{B}\label{ebc11c}.
\end{eqnarray}
Note that now, all terms are linear and the transformation on $\hat{\mathcal L}$ now depends on all parameters: $\left(\epsilon,\xi_A,\eta\right)$.

To realize the algebra, first, let us see that \eqref{ebc10} does not affect the boundary charge, once that it is still integrable:
\begin{equation}
\mathbb{Q}=\int\mathrm{d}\varphi\;\left[\mathcal{L}\epsilon+\mathcal{Q}_A\xi_A+\mathcal{B}\chi\right]=\int\mathrm{d}\varphi\;\left[\hat{\mathcal{L}}\epsilon+\mathcal{Q}_A\xi_A+\mathcal{B}\eta\right].\label{ebc12}
\end{equation}
Decomposing in Fourier modes, we obtain
\begin{eqnarray}
\left[\hat{\mathrm{L}}_{n},\hat{\mathrm{L}}_{m}\right]&=&\left(n-m\right)\hat{\mathrm{L}}_{n+m}+\frac{k}{2}n^{3}\delta_{n+m,0}\label{ebc13a}\\
\left[\hat{\mathrm{L}}_{n},\mathrm{B}_{m}\right]&=&-m\mathrm{B}_{n+m}\label{ebc13b}\\
\left[\mathrm{B}_{n},\mathrm{B}_{m}\right]&=&2kn\delta_{n+m,0}\label{ebc13c}\\
\left[\hat{\mathrm{L}}_{n},\mathrm{Q}_{r}^{A}\right]&=&\left(\frac{n}{2}-r\right)\mathrm{Q}_{n+r}^{A}\label{ebc13d}\\
\left[\mathrm{B}_{n},\mathrm{Q}_{r}^{A}\right]&=&i\epsilon^{AB}\mathrm{Q}_{n+r}^{B}\label{ebc13e}\\
\left\{ \mathrm{Q}_{r}^{A},\mathrm{Q}_{s}^{B}\right\} &=&2\delta^{AB}\hat{\mathrm{L}}_{r+s}+i\epsilon_{AB}\left(r-s\right)\mathrm{B}_{r}+2kr^{2}\delta^{AB}\delta_{r+s,0}\label{ebc13f}.
\end{eqnarray}
The commutator of \eqref{ebc13a} represent the Virasoro algebra with central charge $c=6k$. From the commutators \eqref{ebc13b}, \eqref{ebc13c} we notice that $\mathrm{B}_n$ satisfy the $\mathfrak{u}(1)$ algebra with a positive Kac-Moody level $\kappa=4k$. The relations \eqref{ebc13d}, \eqref{ebc13e}, \eqref{ebc13f} are the algebra of the two supercharges. We obtain the same algebra for the barred sector, as a result, we obtain two copies of the $N=2$ superconformal algebra \cite{extended-susy}, \cite{extended-susy2}.

To complete the $N=2$ superconformal analysis, let us write the temporal components of the auxiliary connection:
\begin{eqnarray}\label{ebc14}
a_t &=& \mu L_1 -\mu'L_0 + \left( \frac{1}{2}\mu''-\frac{2\pi}{k}\mathcal{L}\mu-\frac{\pi}{k}\mathcal{Q}_{A}\nu_{A}\right)L_{-1} + \nu_A G^{\frac{1}{2}}_A
\nonumber\\
&&-\left(\nu_{A}'+\frac{\pi}{k}\mathcal{Q}_{A}\mu+\frac{\pi}{k}\epsilon_{AB}\mathcal{B}\nu_{B}\right)G^{-\frac{1}{2}}_A+\left(\zeta+\frac{\pi}{k}\mathcal B \mu\right)T
\end{eqnarray}
where $\mu\equiv\mu^+$, $\nu_A\equiv\nu_A^{\frac{1}{2}}$ and $\zeta \equiv \sigma - \frac{\pi}{k}\mathcal B \mu$ are the chemical potentials. The equations of motion become:
\begin{eqnarray}
\partial_{t}\hat{\mathcal{L}}&=&-\frac{k}{4\pi}\mu'''+2\hat{\mathcal{L}}\mu'+\hat{\mathcal{L}}'\mu+\frac{3}{2}\mathcal{Q}_{A}\nu_{A}'+\frac{1}{2}\mathcal{Q}_{A}'\nu_{A}+\mathcal{B}\zeta'\label{ebc15a}\\
\partial_{t}\mathcal{Q}_{A}&=&\frac{k}{\pi}\nu_{A}''-2\hat{\mathcal{L}}\nu_{A}+\frac{3}{2}\mathcal{Q}_{A}\mu'+\mathcal{Q}_{A}'\mu+2\epsilon_{AB}\mathcal{B}\nu_{B}'+\epsilon_{AB}\mathcal{B}'\nu_{B}-\epsilon_{AB}\mathcal{Q}_{B}\zeta\label{ebc15b}\\
\partial_{t}\mathcal{B}&=&\frac{k}{\pi}\zeta'+\mathcal{B}\mu'+\mathcal{B}'\mu-\epsilon_{AB}\mathcal{Q}_{A}\nu_{B}\label{ebc15c}.
\end{eqnarray}
In order to make contact with the Brown-Henneaux boundary conditions, we consider $\mu=1$, $\nu_A=\zeta=0$. As a result, the above equations reduce to:
\begin{equation}\label{ebc16}
\partial_{t}\hat{\mathcal{L}}=\hat{\mathcal{L}}',\;\;\;\partial_{t}\mathcal{Q}_{A}=\mathcal{Q}_{A}',\;\;\;\partial_{t}\mathcal{B}=\mathcal{B}'
\end{equation}
and the angular and temporal components of the auxiliary gauge connection coincide: $a_\varphi=a_t$.

\subsection{$N=2$ warped super-conformal boundary conditions}\label{subsec6.2}

This is an extension of the warped super-conformal boundary conditions of section $4.2$. Here we fix all chemical potentials equal to zero and for the angular component of the connection we write
\begin{equation}
a_\varphi = e^{2\phi}L_1 + S,\;\;\;\left[L_{-1},S\right]=0.
\end{equation}
This condition goes beyond the highest weight ansatz \eqref{hw} since the presence of the conformal factor in $L_1$ leads to the fluctuation of the leading order of the metric. Furthermore, this condition fixes the value of $\mathcal L^0,\mathcal Q^{\frac{1}{2}}_A$ to zero and $\mathcal L^+=-\frac{k}{2\pi}e^{2\phi}$. It is convenient to define the following fields:
\begin{equation}\label{ebc17}
\mathcal{L}\equiv e^{2\phi}\mathcal{L}^{-},\;\;\;\mathcal{J}\equiv-\frac{k}{2\pi}\phi',\;\;\;\mathcal{Q}_A\equiv e^{\phi}\mathcal{Q}^{-\frac{1}{2}}_A.
\end{equation}
Therefore, the angular component of the auxiliary gauge connection can be written as:
\begin{equation}\label{ebc18}
a_{\varphi}=e^{2\phi}L_{1}-\frac{2\pi}{k}e^{2\phi}\mathcal{L}L_{-1}-\frac{\pi}{k}e^{-\phi}\mathcal{Q}_{A}G_{-\frac{1}{2}}^{A}+\frac{\pi}{k}\mathcal{B}T
\end{equation}
which is the $N=2$ extension of \eqref{bc19}.

The gauge parameter now depends on five independent components $\epsilon^0, \epsilon,\;\xi_ A$ and $\eta$:
\begin{eqnarray}\label{ebc19}
\lambda&=&b^{-1}\left[ e^{2\phi}\epsilon L_{1}+\epsilon^{0}L_{0}-e^{-2\phi}\left(\frac{1}{2}\epsilon^{0}{}'+\frac{2\pi}{k}\mathcal{L}\epsilon+\frac{\pi}{k}\mathcal{Q}_{A}\xi_{A}\right)L_{-1}+e^{\phi}\xi_{A}G_{A}^{\frac{1}{2}}\right.\nonumber\\
&&\left.-e^{-\phi}\left(\xi_{A}'-\frac{2\pi}{k}\mathcal{J}\xi_{A}+\frac{\pi}{k}\mathcal{Q}_{A}\epsilon+\frac{\pi}{k}\epsilon_{AB}\mathcal{B}\xi_{B}\right)G_{A}^{-\frac{1}{2}}+\left(\eta+\frac{\pi}{k}\mathcal{B}\epsilon\right)T\right]b.
\end{eqnarray}
Note that we have already perform the redefinition of the gauge parameter $\chi\rightarrow\eta$ as we did in the previous section.
The asymptotic symmetry algebra is given by
\begin{eqnarray}
\delta_{\lambda}\mathcal{J}&=&-\frac{k}{4\pi}\epsilon^{0}{}'-\frac{k}{4\pi}\epsilon''+\mathcal{J}\epsilon'+\mathcal{J}'\epsilon\label{ebc20a}\\
\delta_{\lambda}\hat{\mathcal{L}}&=&\frac{k}{4\pi}\epsilon^{0}{}''+\mathcal{J}\epsilon^{0}{}'+2\hat{\mathcal{L}}\epsilon'+\hat{\mathcal{L}}'\epsilon+\frac{3}{2}\mathcal{Q}_{A}\xi_{A}'+\frac{1}{2}\mathcal{Q}_{A}'\xi_{A}+\mathcal{B}\eta'\label{ebc20b}\\
\delta_{\lambda}\mathcal{Q}_{A}&=&\frac{k}{\pi}\xi_{A}''-2\hat{\mathcal N}\xi_{A}+\frac{3}{2}\mathcal{Q}_{A}\epsilon'+\mathcal{Q}_{A}'\epsilon+2\epsilon_{AB}\mathcal{B}\xi_{B}'+\epsilon_{AB}\mathcal{B}'\xi_{B}-\epsilon_{AB}\mathcal{Q}_{B}\eta\label{ebc20c}\\
\delta_{\lambda}\mathcal{B}&=&\frac{k}{\pi}\eta'+\mathcal{B}\epsilon'+\mathcal{B}'\epsilon-\epsilon_{AB}\mathcal{Q}_{A}\xi_{B}\label{ebc20d}
\end{eqnarray}
where
\begin{equation}\label{ebc21}
    \hat{\mathcal N} \equiv \mathcal{L} + \mathcal{J}' + \frac{2\pi}{k}\mathcal{J}^{2} + \frac{\pi}{2k}\mathcal B^2 = \mathcal N + \frac{\pi}{2k}\mathcal B^2.
\end{equation}
It can be easily shown that $\hat{\mathcal N},\; \mathcal Q_A$ and $\mathcal B$ satisfy the transformation of $N=2$ super-conformal algebra.

Now, in this case the boundary charge is also integrable:
\begin{equation}\label{ebc22}
\mathbb{Q}=\int\mathrm{d}\varphi\;\left[\mathcal{J}\epsilon^{0}+\hat{\mathcal{L}}\epsilon+\mathcal{Q}_{A}\xi_{A}+\mathcal{B}\eta\right]
\end{equation}
where $\left(\mathcal J,\hat{\mathcal L}, \mathcal Q_A,\mathcal B \right)$ play now the role of generators of the following asymptotic symmetry:
\begin{eqnarray}
\left[\hat{\mathrm{L}}_{n},\hat{\mathrm{L}}_{m}\right]&=&\left(n-m\right)\hat{\mathrm{L}}_{n+m}\label{ebc23a}\\
\left[\hat{\mathrm{L}}_{n},\mathrm{J}_{m}\right]&=&-m\mathrm{J}_{n+m}+i\frac{k}{2}n^{2}\delta_{n+m,0}\label{ebc23b}\\
\left[\mathrm{J}_{n},\mathrm{J}_{m}\right]&=&-\frac{k}{2}n\delta_{n+m,0}\label{ebc23c}\\
\left[\hat{\mathrm{L}}_{n},\mathrm{Q}_{e}^{A}\right]&=&\left(\frac{n}{2}-r\right)\mathrm{Q}_{n+r}^{A}\label{ebc23d}\\
\left[\hat{\mathrm{L}}_{n},\mathrm{B}_{m}\right]&=&im\mathrm{B}_{n+m}\label{ebc23e}\\
\left[\mathrm{B}_{n},\mathrm{B}_{m}\right]&=&2kn\delta_{n+m,0}\label{ebc23f}\\
\left[\mathrm{B}_{n},\mathrm{Q}_{r}^{A}\right]&=&i\epsilon^{AB}Q_{n+r}^{B}\label{ebc23g}\\
\left\{ \mathrm{Q}_{r}^{A},\mathrm{Q}_{s}^{B}\right\} &=&2\delta^{AB}\hat{\mathrm{N}}_{r+s}+i\epsilon_{AB}\left(r-s\right)\mathrm{B}_{r}+2kr^{2}\delta^{AB}\delta_{r+s,0}.\label{ebc23h}
\end{eqnarray}
This is the warped $N=2$ super-conformal algebra. Note that in this case the asymptotic symmetry algebra is endowed with two $\mathfrak{u}(1)$ generators: $\mathrm{B}_n,\;\mathrm{J}_n$ with different Kac-Moody levels: $\kappa_{B}=4k$ and $\kappa_{J}=-k$.

\subsection{Avery-Poojary-Suryanarayana boundary conditions for $\mathcal N=2$ supergravity}

In this case we work with the $\mathrm{Osp}\left(2\mid2\right)\otimes \mathrm{SL}(2)$ Chern-Simons gravity. We choose the extended supersymmetric Brown-Henneaux boundary conditions for the $\Gamma$-sector. On the $\bar{\Gamma}$-sector, we only have $\mathcal T^I\equiv\bar{\mathcal L}^I$ fields varying (All fermions and $\mathrm{SO}(2)$ fields are set to zero). In this case we perform the Sugawara shift
\begin{equation}\label{ebc24}
    \hat{\mathcal{L}}\equiv\mathcal{L}^{-}+\frac{2\pi}{k}\left(\mathcal{T}^{0}\mathcal{T}^{0}-\mathcal{T}^{+}\mathcal{T}^{-}+\frac{1}{4}\mathcal{B}^{2}\right)=\hat{\mathcal{L}}^{-}+\frac{2\pi}{k}\left(\mathcal{T}^{0}\mathcal{T}^{0}-\mathcal{T}^{+}\mathcal{T}^{-}\right)
\end{equation}
which allows us to obtain the semidirect sum of the $N=2$ super-conformal algebra and the $\mathrm{sl}\left(2\right)_k$ algebra.

\section{Final Remarks:}\label{sec7}

In this paper we used the Chern-Simons formulation to analyse the asymptotic symmetries of the $\mathcal{N}=\left(1,1\right)$ and $\mathcal{N}=\left(2,2\right)$ $\mathrm{AdS}_3$ supergravity. Using the same procedure of \cite{Grumiller3D} we found that for the most general boundary conditions the asymptotic symmetry algebras are two copies of the $\mathfrak{osp}(1\mid 2)_k$ and $\mathfrak{osp}(2\mid 2)_k$ algebra, respectively. However, the results can be easily applied to $\mathcal N=1$ or $\mathcal N=2$ supergravity: For the $\Gamma$-sector we have the $\mathfrak{osp}(1\mid 2)_k$ (or $\mathfrak{osp}(2\mid 2)_k$) algebra, while for the $\bar\Gamma$-sector -which only have bosonic charges- the $\mathfrak{sl}(2)_k$ algebra.

We then restrict to particular cases, fixing some chemical potentials and auxiliary fields while allowing others  to vary. Not any arbitrary choice lead us to an asymptotic algebra. In particular, here we have shown that for the supersymmetric Brown-Henneaux boundary conditions we obtain two copies of the $N=1$ and $N=2$ super-conformal algebras, these results are useful to test the validity of the formalism. For the supersymmetric Troessaert boundary conditions we obtain two copies of $N=1$ and $N=2$ warped super-conformal algebras. As far as we know, these results are novels in the context of three-dimensional gravity. However, in \cite{AdS2SUSY} was shown a single copy of the warped super-conformal algebra as a result of the asymptotic analysis of the two-dimensional Jackiw-Teitelboim supergravity. In \cite{Poojary}, it was shown a supersymmetric extension of the Avery-Poojary-Suryanarayana boundary conditions for (extended) supergravity. In this work we decide to restrict to $\mathcal  N=1$  and $\mathcal N=2$ supergravity and, along with the Sugawara shifts in equations \eqref{bc29} and \eqref{ebc24}, show explicitly that the asymptotic symmetry algebra is a semidirect sum of the $N=1$ (or $N=2$) super-conformal algebra and the $\mathfrak{sl}(2)_k$ algebra. In total, we have four boundary conditions for (extended) supergravity, this may looks like a more restricted set of boundary conditions that the one obtained in the pure bosonic case \cite{Grumiller3D}. However, equations  \eqref{bc06a}-\eqref{bc06c} and \eqref{ebc04a}-\eqref{ebc04d} can be used to obtain new boundary conditions: For example, if we want to work with the Avery-Poojary-Suryanarayana boundary conditions for $\mathcal N = \left(1,1\right)$ gravity we must allow all bosonic and fermionic charges in the $\bar{\Gamma}$ sector to vary. Then, we need to perform a new Sugawara shift  such that the asymptotic symmetry algebra is the semidirect sum of a super-Virasoro and the $\mathfrak{osp}\left(1\mid 2\right)_k$ algebra. Another boundary conditions that can be explored are the ones obtained in the context of two-dimensional Jackiw-Teitelboim gravity \cite{AdS2SUSY} where we only set to zero one fermion charge and leave all bosons to vary. 

It is also important to notice that for the three particular cases of boundary conditions for extended supergravity, we need to perform a shift (see equations \eqref{ebc10}, \eqref{ebc21} and \eqref{ebc24}) in order to obtain a linear asymptotic symmetry algebra. This shift was already performed in \cite{Henneaux-SUSY} for the Brown-Henneaux boundary conditions for extended supergravity and now we have generalized that result for other boundary conditions. Furthermore, for the $N=2$ warped super-conformal case we obtain two $\mathfrak{u}(1)$ currents, one of them with a negative Kac-Moody level that may lead to nonunitary representations \cite{WCFT01}, \cite{WCFT02}.

As future perspectives of this work, it would be interesting to explore boundary conditions related to supersymmetric $\mathrm{KdV}$ hierarchy \cite{sKdVa}, \cite{sKdVb}, \cite{sKdVc} following a similar approach as in \cite{Troncoso}, \cite{Gonzales}.

\section*{Acknowledgements}

We thank D. Vassilevich for reading the manuscript and helpful comments. We also thanks M. C\'ardenas, O. Fuentealba, H. A. Gonz\'alez and D. Grumiller for their collaboration on two-dimensional supergravity. CEV was partially supported by CAPES.

\section*{Appendix: The $\mathfrak{osp}\left(N\mid2\right)$ super-algebra}\label{app}

A super-algebra is denoted by: $\mathfrak{g}=\mathfrak{g}_{0}\oplus\mathfrak{g}_{1}$, where $\mathrm{g}_{0}$ and $\mathrm{g}_{1}$ are the even and odd part of the super-algebra, respectively. In order to describe (extended) supergravity, it is necessary to work with the $\mathfrak{g}=\mathfrak{osp}(N\mid 2)$ algebra. The even part of this algebra is $\mathfrak{g}_0=\mathfrak{sl}(2)\oplus\mathfrak{so}(N)$, with generators $L_I\;(I=0,\pm1)$ and $T_{AB}=-T_{BA}\; (A,B=1,2,...,N)$. The odd part of the super-algebra $\mathfrak{g}_1$ has generators $G^\alpha_A,\;\left(\alpha=\pm\frac{1}{2}\right)$. These generators satisfy the following algebra
\begin{eqnarray}
\left[L_I,L_J\right] &=& (I-J) L_{I+J}\\
\left[L_{I},G^A_{\alpha} \right]	&=&	\left(\frac{I}{2}-\alpha\right)G^A_{I+\alpha}\\
\left[ G^A_\alpha, T^{AB} \right] &=& \delta^{AB}G^C_\alpha - \delta^{AC}G^B_\alpha\\
\left[ T^{AB}, T^{CD} \right] &=& \delta^{AD}T^{BC}+\delta^{BC}T^{AD}-\delta^{AC}T^{BD}-\delta^{BD}T^{AC} \\
\left\{ G^A_{\alpha},G^B_{\beta} \right\} &=& -2L_{\alpha+\beta}\delta^{AB}-\left(\alpha-\beta\right)T^{AB}.
\end{eqnarray}
The supertrace $\left(\mathrm{Str}\right)$ is a bilinear form and its nonvanishing components are denoted by: $\mathrm{Str}(L_I L_J)=\kappa_{IJ}$, $\mathrm{Str}(G^A_\alpha G^B_\beta) = \kappa^{AB}_{\alpha\beta}$ and $\mathrm{Str}(T^{AB}T^{CD}) = \kappa^{ABCD}$, such that
\begin{eqnarray}
&&\kappa_{00}=1/2,\;\;\;\; \kappa_{+-} = \kappa_{-+}=-1,\;\;\;\; \kappa^{AB}_{\frac{1}{2},-\frac{1}{2}}=-\kappa^{AB}_{-\frac{1}{2},\frac{1}{2}}=2\delta^{AB},\label{ap01a}\\
&&\kappa^{ABCD}=2\left(\delta^{AC}\delta^{DB}-\delta^{AD}\delta^{CB}\right)=\kappa^{CDAB}=-\kappa^{BACD}.\label{ap01b}
\end{eqnarray}

For $N=1$, there are no $T^{AB}$ generators and we can drop the $\mathfrak{so}(N)$ indices. Therefore, the $\mathfrak{osp}\left(1\mid 2\right)$ algebra is given by:
\begin{eqnarray}
\left[L_I,L_J\right] = (I-J) L_{I+J},\;\;\;\left[L_{I},G_{\alpha} \right]	=	 \left(\frac{I}{2}-\alpha\right)G_{I+\alpha},\;\;\;
\left\{ G_{\alpha},G_{\beta} \right\}  = -2L_{\alpha+\beta}.\label{ap02}
\end{eqnarray}

For $N=2$, due to the antisymmetry of the $\mathfrak{so}\left(2\right)$ generators we can write $T^{AB}=\epsilon^{AB}T$. Therefore, the algebra reduces to
\begin{eqnarray}
\left[L_{I},L_{J}\right]=	\left(I-J\right)L_{I+J},\;\;\;\;&&
\left[L_{I},G_{\alpha}^{A}\right]	=	\left(\frac{I}{2}-\alpha\right)G_{I+\alpha}^{A},\label{ap03a}\\
\left[G_{\alpha}^{A},T\right]	=	\epsilon^{AB}G_{\alpha}^{B},\;\;\;\;&&
\left\{ G_{\alpha}^{A},G_{\beta}^{B}\right\} =	 -2L_{\alpha+\beta}\delta^{AB}-\left(\alpha-\beta\right)\epsilon^{AB}T.\label{ap03b}
\end{eqnarray}
As a direct consequence of \eqref{ap01b}, we have $\kappa=\mathrm{Str}(T^2)=2$.

\end{document}